\documentclass[twocolumn,showpacs,amsmath,prb,superscriptaddress]{revtex4-1}
\usepackage{color}
\usepackage{graphicx}
\usepackage{epsfig,psfrag}
\usepackage{amsmath}
\usepackage{amssymb}
\usepackage{amsbsy}
\usepackage{amsthm}
\usepackage{amsfonts}
\usepackage{wasysym}
\usepackage{bbm}
\usepackage{tabularx}
\usepackage{euscript}
\usepackage{enumerate}
\usepackage{amsfonts}
\usepackage{exscale}
\usepackage{bbold}
\usepackage{float}
\usepackage{slashed}
\usepackage{subfigure}
\usepackage{comment}
\usepackage[colorlinks,citecolor=blue]{hyperref}

\newcommand{\bea}{\begin{eqnarray}}
\newcommand{\eea}{\end{eqnarray}}
\newcommand{\ba}{\begin{array}}
\newcommand{\ea}{\end{array}}

\def\bea{\begin{eqnarray}}
\def\eea{\end{eqnarray}}

\def\nn{\nonumber}

\begin{document}

\title{${\bf 2k_F}$ Density Wave Instability of Composite Fermi Liquid}

\author{Shao-Kai Jian}
\affiliation{Condensed Matter Theory Center, Department of Physics, University of Maryland, College Park, Maryland 20742, USA}

\author{Zheng Zhu}
\email[\href{mailto:zhuzhengphysics@gmail.com}{zhuzhengphysics@gmail.com}]{}
\affiliation{Kavli Institute for Theoretical Sciences, University of Chinese Academy of Sciences, Beijing 100190, China}
\affiliation{Department of Physics, Harvard University, Cambridge, Massachusetts 02138, USA}

\date{\today}

\begin{abstract}
We investigate the $2k_F$ density-wave instability of non-Fermi liquid states by combining exact diagonalization with renormalization group analysis. At the half-filled zeroth Landau level, we study the fate of the composite Fermi liquid in the presence of the mass anisotropy and mixed Landau level form factors. These two experimentally accessible knobs trigger a phase transition towards a unidirectional charge-density-wave state with a wavevector equal to $2k_F$ of the composite Fermi liquid. Based on exact diagonalization, we identify such a transition by examining both the energy spectra and the static structure factor of charge density-density correlations. Moreover, the renormalization group analysis reveals that gauge fluctuations render the non-Fermi liquid state unstable against density-wave orders, consistent with numerical observations. Possible experimental probes of the density-wave instability are also discussed.
\end{abstract}

\maketitle

\section{Introduction}

Non-Fermi liquids (NFLs) are among the most exotic quantum states in condensed matters. One class of NFL states is realized at quantum critical points (QCPs)~\cite{Hertz1976, Wolfle2007, SachdevBook, Lee2018} with gapless collective mode. The most well-known example is the strange metal, which has been intensively investigated after the discoveries of high-temperature superconductors~\cite{Muller1986} and heavy-fermion materials~\cite{Coleman2007}. More recently, Moir\'{e} materials such as the twisted bilayer graphene~\cite{Cao2018} have created new excitement.
Instead of appearing at QCPs, the NFL state can also arise as a stable phase at zero temperature.
A prominent example is the two-dimensional (2D) electrons under a strong magnetic field: when the zeroth Landau level (LL) is half-filled, it becomes a fractionalized gapless state~\cite{Halperin1993,Son2015} with a large Fermi surface formed by composite fermions (CFs)~\cite{Jain1989,JainBook}.

Fathoming the instabilities of NFL is the very essence of understanding various phenomena  in strongly correlated systems. For example, the high transition temperature and the complex orders of the high-temperature superconductors are all believed to result from a NFL mother state~\cite{Keimer2015,Lee2008,Kivelson2003,Lee2006}.
Theoretically a stable and controllable platform is crucial and urgently needed for investigating the intriguing properties of the NFL states.
In particular, the compressible NFL state at the half-filled LL is well established both experimentally~\cite{Willett1997, Du1993, Du1994} and numerically~\cite{Motrunich2016, WangJie, Jain2015}, which provides a promising platform.
More importantly, the physical setup also comes with various tuning knobs such as the magnetic field, the geometry, and the number of components including layers, subbands, spins and/or valleys.
With these knobs, plenty of states adjacent to the composite Fermi liquid (CFL) are discovered, consequently revealing various instabilities of CFL.
For instance, the Cooper instability~\cite{Bonesteel1999,Senthil2015} leads to the $p+i p$ paired Moore-Read (MR) state~\cite{Moore1991, Greiter1992, Read2000} (we briefly review it in Appendix~\ref{append:pair});
the Pomeranchuk instability~\cite{Fradkin2014, Kim2018} results in nematic quantum Hall states~\cite{West1999, Du1999}; the Stoner instability of CFL gives rise to spin or valley polarizations~\cite{JainStoner,Balram2015,ZZ2018}; and the instability towards the Halperin 331-state~\cite{Halperin1983, SDS2010} in quantum Hall bilayers.

In this paper, we propose one mechanism to reap yet another instability of CFL: the $2k_F$ density-wave instability,which is of equal importance to the previously discovered CFL instabilities and is likely to exhibit distinct physics from ordinary Fermi liquids~\cite{Altshuler1995, Metzner2018, Punk2019}.
Based on an exact diagonalization (ED) and renormalization group (RG) analysis, we propose one possible mechanism to trigger the density-wave instability of CFL on half filled LLs: tuning the interactions via the mixed LL form factors from an anisotropic CFL state. We  demonstrate such an instability numerically
and reveal the underlying mechanism by RG analysis. We find the density-wave instability would be dominant over the pairing instability via increasing the gauge fluctuations, which can be achieved by breaking the rotational symmetry. Importantly, the mixed form factor is experimentally accessible in Dirac materials, e.g., in bilayer graphene, by tuning the interlayer electric bias and the magnetic field~\cite{Papic2011,Young2017a,Young2017b,ZZ2019},  rendering it possible to examine our findings.

%----------------------------------------------------------------------------------------------------------------------------------------------------------------------------------------
\begin{figure*}[tbp]
\begin{center}
\includegraphics[width=5.5cm]{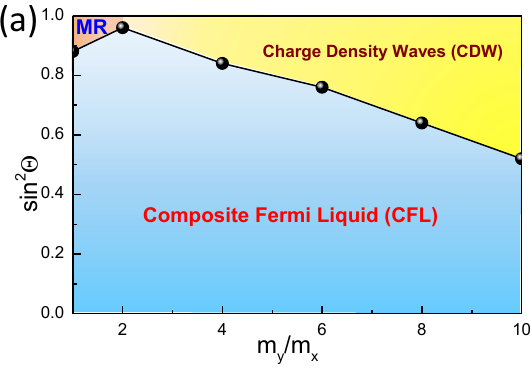} \qquad \qquad 
\includegraphics[width=5.5cm]{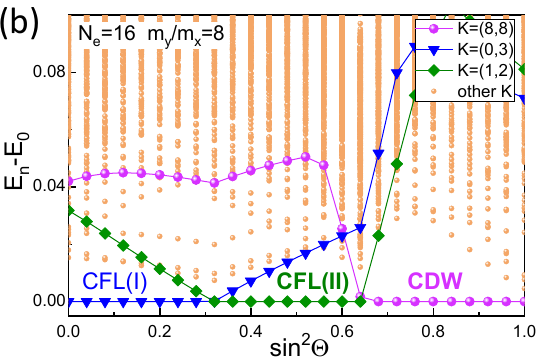} \\
\includegraphics[width=6cm]{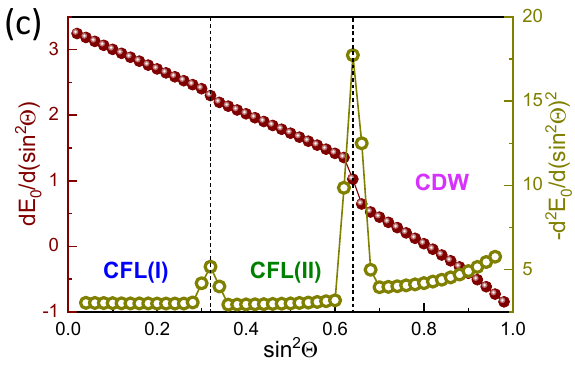} \qquad 
\includegraphics[width=5.7cm]{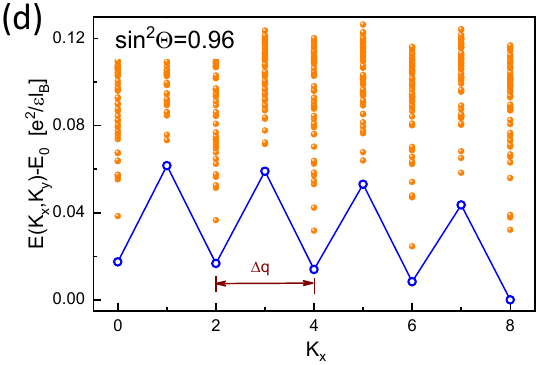} \\
\end{center}
\par
\renewcommand{\figurename}{Fig.}
\caption{(Color online) {The phase diagram and the energy spectra.} Depending on the mass anisotropy $m_y/m_x$, we identify the pairing instability and density-wave instability of CFL when tuning the interaction via $\text{sin}^2\Theta$, and the corresponding phase diagram is shown in panel (a). For a fixed mass ratio, e.g., $m_y/m_x=8$ in panel (b-d), the phase boundary is consistently identified from the evolution of energy spectra with $\text{sin}^2\Theta$  (b)  and the derivatives of the ground-state energy (c). In the charge density wave phase, the energy spectra along momentum $K_x$ exhibits the quasidegenerate states that differ by a momentum $\Delta q$ (d).  Here, we consider a half-filled Landau level with $N_e=16$ electrons.}
\label{Fig1}
\end{figure*}
%----------------------------------------------------------------------------------------------------------------------------------------------------------------------------------------
\begin{figure*}[tbp]
\begin{center}
	\subfigure[]{
	\includegraphics[width=5.5cm]{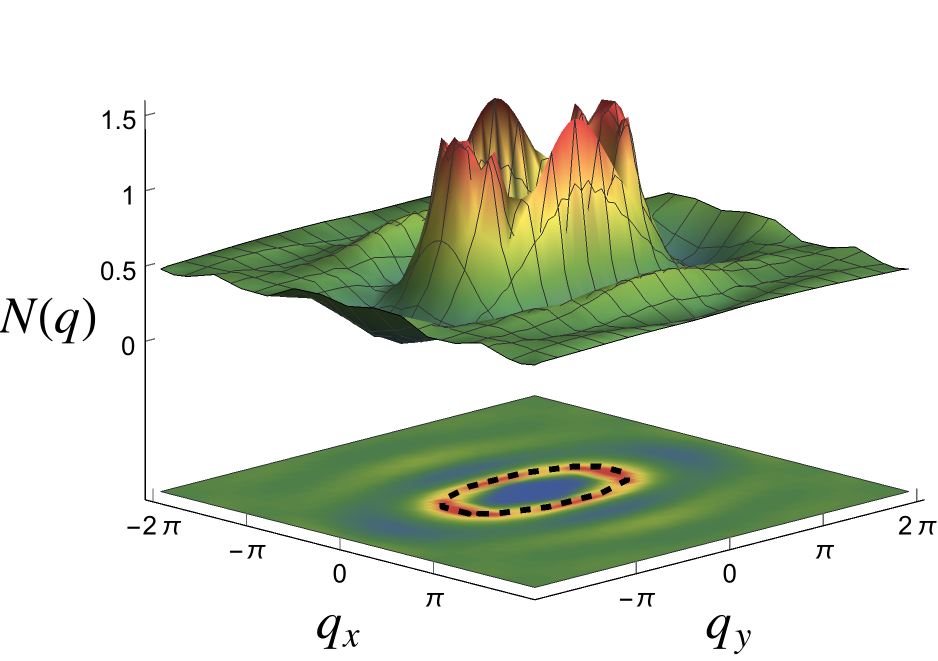}} \qquad 
	\subfigure[]{
	\includegraphics[width=5.5cm]{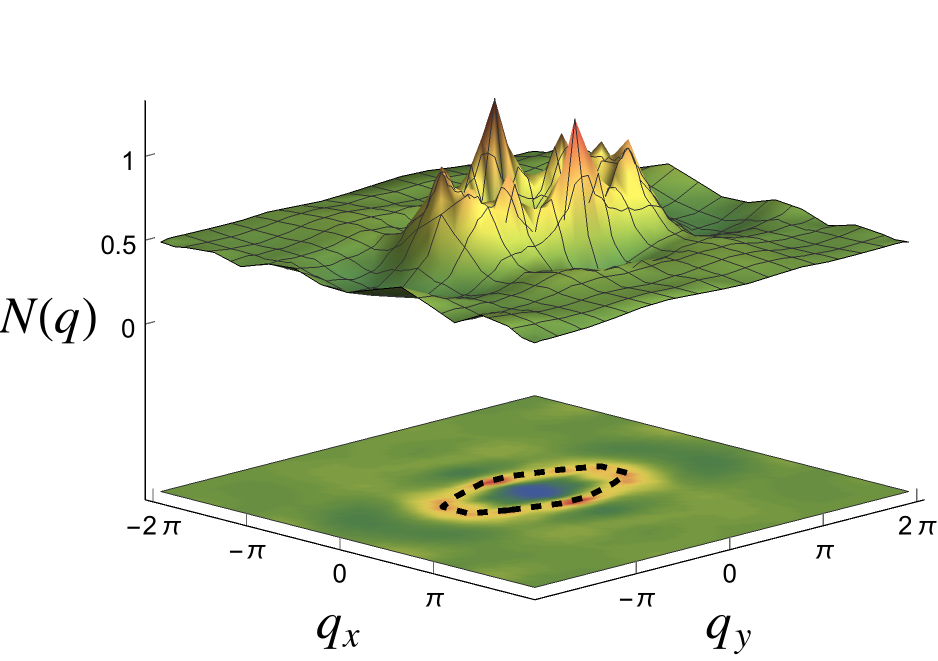}} \\
	\subfigure[]{
	\includegraphics[width=5.5cm]{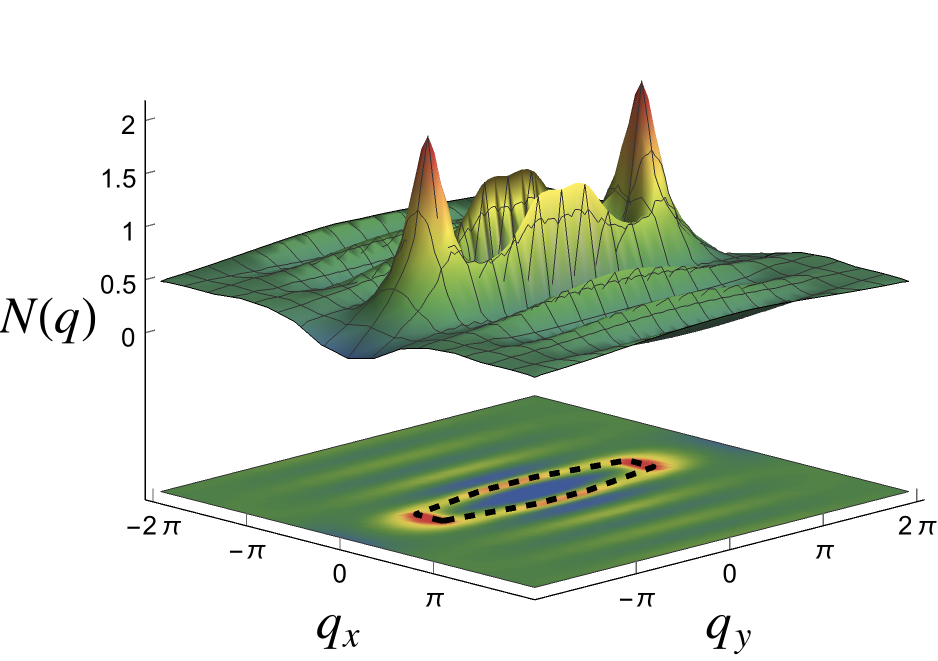}} \qquad
	\subfigure[]{
	\includegraphics[width=5.5cm]{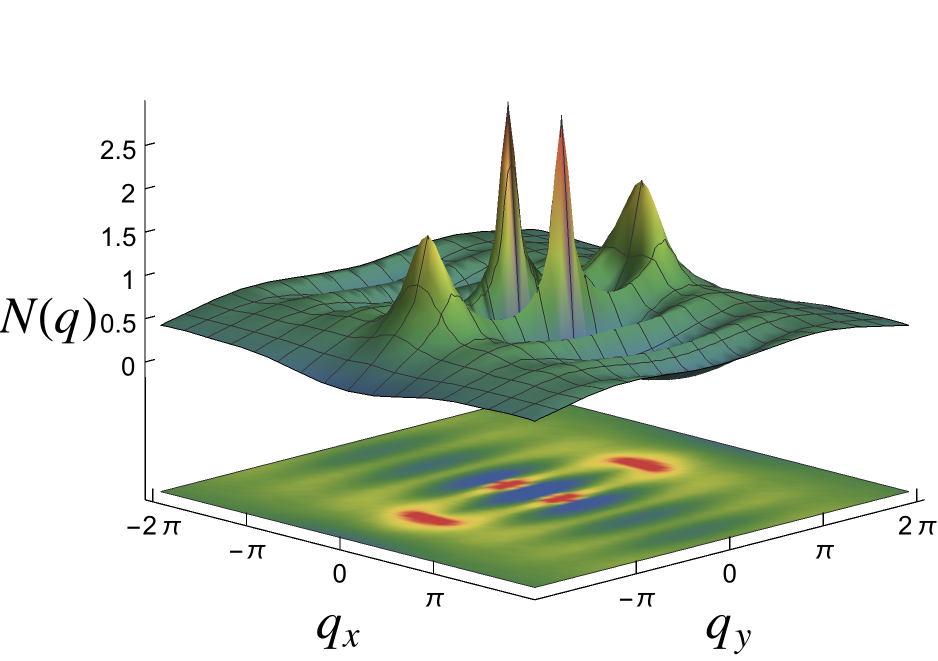}}
\end{center}
\par
\renewcommand{\figurename}{Fig.}
\caption{(Color online) {The static structure factors $N(\mathbf{q})$.} The nature of the different phases in Fig.~\ref{Fig1}(b-c) can be identified from the static structure factor $N(\mathbf{q})$ of the density-density correlation.  Panels (a-d) show $N(\mathbf{q})$ in the CFL phase with  $\text{sin}^2\Theta=0.16$ (a) and  $\text{sin}^2\Theta=0.48$ (b) , as well as $N(\mathbf{q})$ in the charge density wave phase with $\text{sin}^2\Theta=0.68$ (c) and $\text{sin}^2\Theta=0.96$ (d).The dashed line in (a), (b) and (c) indicates the Fermi surface of CFs. Note a factor of two between the scattering momentum and the momentum of CFs. Here, we consider a half-filled Landau level with $N_e=16$ electrons and mass ratio $m_y/m_x=8$. }
\label{Fig:Nq}
\end{figure*}

\section{Numerical Setup and Results}

We consider 2D electrons on a torus with a strongly perpendicular magnetic
field piercing through its surface. The Hamiltonian is given by
 \begin{equation}\label {Ham}
H=\frac{1}{2A}\sum_{{\mathbf{q}}}V({\mathbf q})F({\mathbf {q}})F({\mathbf {-q}}):\rho^\dagger({\bf q})\rho({\bf q}):,~
  \end{equation}
where $V(\mathbf{q})$ is the Fourier transform of the un-projected Coulomb interaction, $F({\mathbf q})$ denotes the density form factor introduced by projection, $\rho({\mathbf q})$ is the guiding center density operators, and $A$ represents the area of the 2D plane.
Below we consider the mixed form factors $F(\mathbf{q})= \text{cos}^2\Theta F_0(\mathbf{q}_m)+\text{sin}^2\Theta F_1(\mathbf{q}_m)$ to tune the interactions~\cite{Papic2011,Young2017a,Young2017b,ZZ2019}, where $F_{0,1}(\mathbf{q}_m)=\exp(-\mathbf{q}^2_m/4) L_{0,1}[\mathbf{q}^2_m/2]$ are the form factors for $n=0$ and $n=1$ Galilean LLs, respectively. $L_n(x)$ is the Laguerre polynomial. The anisotropic CFL can be achieved by introducing the mass anisotropy, where $\mathbf{q}^2_m=g^{ab}_mq_aq_b$ includes the metric $g_m =\text{diag}[\sqrt {m_y/m_x},\sqrt {m_x/m_y}]$ derived from the band mass tensor.
In the isotropic limit (i.e., $m_y=m_x$), the CFL and MR states are stabilized at $\text{sin}^2\Theta=0$~\cite{Jain1989,Halperin1993} and $\text{sin}^2\Theta=1$~\cite{Moore1991, Greiter1992, Read2000}, respectively. The corresponding pairing instability in this limit, such as tuning  $\text{sin}^2\Theta$,  has been  theoretically confirmed~\cite{HR2000,Moller,Papic2012}, though the nature of this transition is still controversial~\cite{Bonesteel1999,Moore1991, Greiter1992, Read2000,Senthil2015}. The mass anisotropy explicitly breaks the spatially rotational symmetry~\cite{Haldane2011,Mulligan2010,Yang2012,Qiu2012,Papic2013,Yang2013,Haldane2016,Balram2016,Ippoliti2017,ZZ2017}, concealing another factor to trigger the instability of CFL.
Previous studies have demonstrated that CFL is remarkably robust against mass anisotropy when $\text{sin}^2\Theta=0$~\cite{Ippoliti2017}, while the MR state is fragile against mass anisotropy and finally translates to a stripe state~\cite{Rezayi1999} when $\text{sin}^2\Theta=1$~\cite{ZZ2017}. Then it is natural to investigate the possible density-wave instability of CFL by tuning the interactions via $\text{sin}^2\Theta$ from an anisotropic CFL state at $\text{sin}^2\Theta=0$.  Below we will detect such a possibility by solving the Hamiltonian by ED~\cite{Haldane1985}.

Our numerical results are depicted in the phase diagram shown in Fig.~\ref{Fig1}(a).  In the isotropic limit, we have confirmed the pairing instability of CFL when tuning the interaction via $\text{sin}^2\Theta$, consistent with previous studies. In the presence of mass anisotropy, we find the pairing instability only survives in a small regime in the phase space, and instead, the density-wave instability becomes the dominant instability of CFL after rotational symmetry breaking, which can be triggered more easily by increasing the mass anisotropy [see Fig.~\ref{Fig1}(a)].

The phase boundaries in Fig.~\ref{Fig1}(a) are identified from both the energy spectra and the derivatives of the ground-state energy. Figure~\ref{Fig1}(b) shows an example of the energy spectra as a function of  $\text{sin}^2\Theta$ for an $N_e=16$ system with $m_y/m_x=8$. Further results of $N_e=12, 14$ are given in Appendix~\ref{append:size}. The CFL state is robust up to $\text{sin}^2\Theta\approx0.64$ upon tuning the interaction, which can be further confirmed from the derivatives of the ground-state energy in Fig.~\ref{Fig1}(c). The energy gap in the spectra of CFL is induced by the shell-filling effect on a finite sized system, which can be identified by comparing the quantum number of the ground state obtained by ED and the CFL wavefunctions on a torus~\cite{RR1994,Haldane1985,HR2000,ZZ2018,Read1994}. The energy level crossing near $\text{sin}^2\Theta\approx0.32$ represents the change of the CFL ground-state momentum sectors, in contrast to the phase transitions around  $\text{sin}^2\Theta\approx0.64$. We further confirm the nature of these phases by studying the static structure factor $N(\mathbf{q})$ of the density-density correlation, $N(\mathbf{q})=\frac{1}{N}\langle {\rho_\mathbf{q}\rho _{-\mathbf{q}}} \rangle  = \frac{1}{N}\sum_{i,j} {\langle {{e^{i{\mathbf{q}} \cdot {\mathbf{R}_i}}}{e^{ - i{\mathbf{q}} \cdot {\mathbf{R}_j}}}}\rangle }$,
 where ${\rho _{\bf{q}}} = \sum\nolimits_{i = 1}^N {{e^{i{\bf{q}} \cdot {{\bf{R}}_i}}}} $ is the Fourier transform of the guiding center density. As shown in Fig.~\ref{Fig:Nq}(a-b) for $\text{sin}^2\Theta\lesssim 0.64$, $N(\mathbf{q})$ exhibits a strong $2k_F$ scattering feature induced by the scattering among CFs close to the Fermi surface. At $\text{sin}^2\Theta > 0.64$, there are two sharp peaks in $N(\mathbf{q})$ in the same direction, which can be regarded as the hallmark of charge ordering with the wave vector determined by the position of the peaks. Here, $N(\mathbf{q})$ displays a stripe feature.

Further increasing $\sin^2 \Theta \gtrsim 0.88 $, the peaks rotate from $(q_x,q_y)=(0, \pm q^*)$ to $(q_x,q_y)=(\pm q^{**},0)$ as shown in Fig.~\ref{Fig:Nq}(c-d).
Here, the wave vector $\pm q^{**}$ also can be identified from the low energy spectra of such resulting phase [see Fig.~\ref{Fig1}(d)], where there is no recognizable gap separating the ground-state manifold from the excited states, and instead, the energy spectra displays a conspicuous set of quasi-degenerate states which differ by momentum $\Delta q$ and satisfy $\Delta q=\pm q^{**}$. The line connecting the lowest energy states in each momentum sector has a zigzag structure as shown in Fig.~\ref{Fig1}~(d), which only appears in the energy spectra in one momentum direction, implying a unidirectional charge density wave state.

\section{RG analysis from CFL} 
As the Fermi surface and its instability are indicated in Fig.~\ref{Fig:Nq}, it is natural to understand it within the context of the Halperin-Lee-Read (HLR) theory~\cite{Halperin1993}.
Because the instability is peaked at two antipodal Fermi points [see Fig~\ref{Fig:Nq}~(c)], we use the patch theory~(cf. Chapter 18 of \cite{SachdevBook} for a review) to analyze the competing fluctuations.
The composite-Fermi surface is approximated by two patches~\cite{Senthil2010, Kim2017} near the antipodal Fermi points:
$\mathcal{H}_f = - i s v_F \partial_x - \frac1{2K} \partial_y^2$, where $s= \pm$ denotes the two patches, $\psi_s$ refers to CF, and $x, y$ are the normal and tangent directions of the Fermi surface, respectively.
$v_F$ and $K$ capture the CF Fermi velocity and the curvature of the patch.
Including the gauge field fluctuation, the total effective action is given by $S = S_f + S_a + S_{\text{int}}$,
\bea
	S_f &=& \sum_s \int d^3 x \psi_s^\dag (\partial_\tau + \mathcal H_f ) \psi_s,  \\
	S_a &=& \int_k |k_y|^{1+\epsilon} |a(k)|^2, \\
	S_{\text{int}} &=& \sum_s  s e \int d^3 x a(x) \psi_s^\dag(x) \psi_s(x),
\eea
where $ \int_k \equiv \int \frac{d^3k}{(2\pi)^3}$,  $a(x)$ is the emergent gauge field, and $e$ is the Yukawa coupling between the fermion and gauge boson.
$\epsilon$ is the expansion parameter, and $\epsilon = 0$ corresponds to the long-range Coulomb interaction~\cite{Senthil2010, Kim2017}.

The patch theory is an effective description in the range $ |k_x|, k_y^2 < \Lambda$ (note that $k_x$ and $k_y$ scale differently). We address the IR properties of the theory by integrating out the high energy mode, $ \Lambda e^{-l} < k_y^2 < \Lambda$ to generate RG equations, where $l>0$ is the running parameter. There is no renormalization to the boson propagator because it is nonlocal. The rationale for using a nonlocal bare kinetic term for the gauge boson lies in the fact that the boson kinetic potential does not receive corrections up to three-loop~\cite{Sachdev2010}. Taking into account of the fermion self-energy
$\Sigma_s(p) = -e^2 \int_k D(k) G_s(k+p)\approx - \frac{i e^2}{4 \pi^2 v_F} p_0$,
the RG equation reads (see Appendix~\ref{append:RG})
\bea
\frac{d g}{dl} = \frac{\epsilon}2 g - \frac{g^2}{4},
\eea
where $g \equiv \frac{e^2}{\pi^2 v_F \Lambda^{\epsilon/2}}$ captures the effective Yukawa coupling. The presence of a nontrivial stable fixed point $g^*=2 \epsilon$ corresponds to the NFL interacting strongly with the gauge field.

\begin{figure*}
	\centering
	\subfigure[]{\label{forward1}
		\includegraphics[height=2cm]{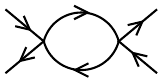}} \qquad
	\subfigure[]{\label{forward2}
		\includegraphics[height=2cm]{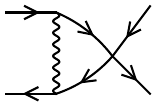}} \qquad
	\subfigure[]{\label{forward3}
		\includegraphics[height=2cm]{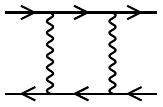}}
	\caption{(Color online)  {The corrections to short-ranged four-fermion interactions within the patch theory}. Panel~(a) denotes the correction from the four-fermion interaction, and panels (b-c) denote corrections from the gauge fluctuations.}
\end{figure*}

Next, we analyze the density-wave instability in the CFL.
Because we are interested in the $2k_F$ instability connecting antipodal Fermi points,
we can consider the scattering processes within the patch theory, namely, $S = S_f + S_a + S_{\text{int}} + S_4$,  $S_4 = U \int d^3 x \psi_+^\dag \psi_+ \psi_-^\dag \psi_- $.
In the patch theory, the four-body interaction is irrelevant, which is consistent with the fact that the forward-scattering process does not affect the existence of the Fermi surface~\cite{Shankar1994}, and the perturbative calculation should be valid. As indicated in Fig.~\ref{forward1}, the renormalization to the four-body interaction reads
\bea
	\Gamma^{(a)}_4 &=& - 2U^2 \int_k G_+(k) G_-(k) 
	\approx \frac{\alpha_0}{\sqrt 2 \pi^2} \frac{\sqrt{\Lambda K} U^2}{v_F} l , \nn
\eea
where $\alpha_0 \equiv \Gamma(0,1)\approx 0.219$, and $\Gamma(n,x) \equiv \int_x^\infty dt t^{n-1} e^{-t} $ is the incomplete Gamma function.
Without gauge fluctuation, the RG equation of dimensionless coupling constant $u \equiv \frac{ \sqrt{K\Lambda}}{\pi^2 v_F} U$ is
\bea
	\frac{du}{dl}= - \frac{u}2+ \sqrt2 \alpha_0 u^2,
\eea
which shows that an instability only occurs at finite interaction strength.
When $u$ is large enough, i.e., $u>\frac{1}{2\sqrt{2}\alpha_0}$, it develops a wave-density instability with the $2k_F$ order parameter $\phi = \psi_+^\dag \psi_-$.

\begin{figure*}[tbp]
	\centering
	\subfigure[]{\label{u-g1}
		\includegraphics[height=4cm]{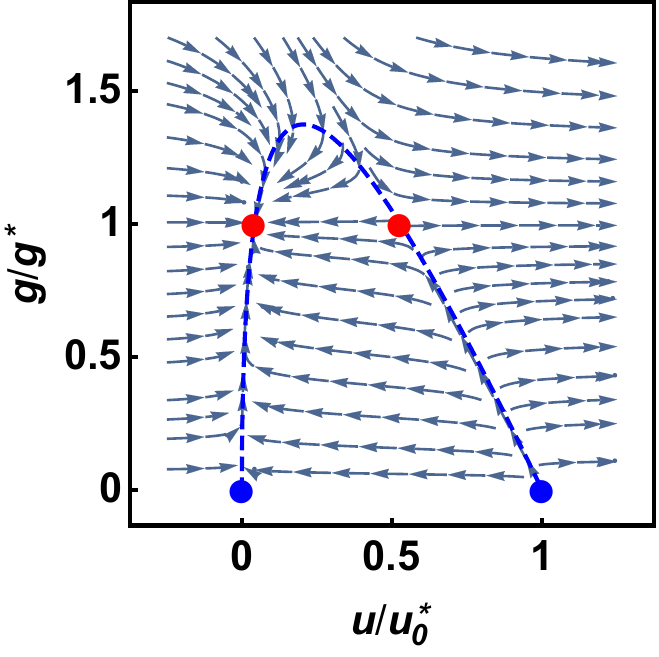}} \qquad
	\subfigure[]{\label{u-g2}
		\includegraphics[height=4cm]{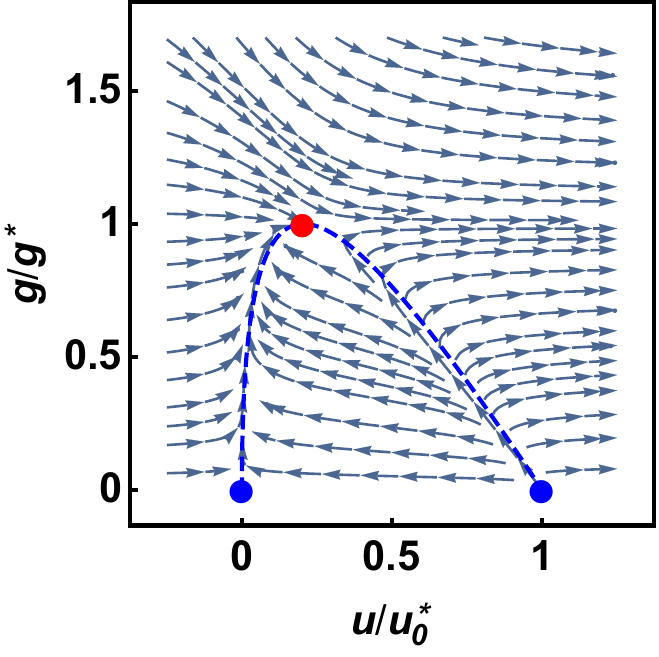}} \qquad
	\subfigure[]{\label{u-g3}
		\includegraphics[height=4cm]{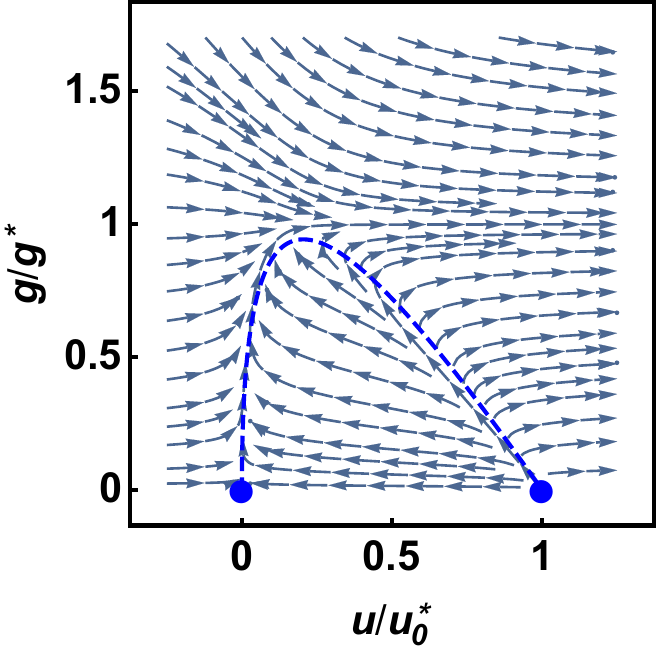}}
	\caption{(Color online) {The RG flow diagrams of $(u, g)$ at different $\epsilon$. }The blue points show the Gaussian fixed point and stripe transition point without gauge fluctuations. Red points show the NFL fixed point and stripe transition point in the presence of gauge fluctuations. Fig.~\ref{u-g1} shows four fixed points when $\epsilon<\epsilon_c$. Figs.~\ref{u-g2},~\ref{u-g3} show the RG flow for $\epsilon=\epsilon_c$, $\epsilon>\epsilon_c$, respectively. The dashed line indicates the trajectory of two nontrivial fixed points in the presence of gauge fluctuations. After their collision, the fixed points become imaginary values, and disappeare from the flow diagram.}
\end{figure*}

Now we consider the effect of gauge fluctuations. As shown in Figs.~\ref{forward2},~\ref{forward3}, the corrections from gauge fluctuation read
\bea
	\Gamma^{(b)}_4 &=& - \frac{2e^2}{3} \int_k G_R(k) G_L(k) D(k)
	\approx \frac{\alpha_0}{3\pi^2} \frac{e^2 u}{v_F} l, \nn \\
	\Gamma^{(c)}_4 &= & - \frac{e^4}{N^2} \int_k G_R(k) G_L(k) D^2(k) \approx \frac{\alpha_0}{2\sqrt2 \pi^2} \frac{e^4}{\sqrt{K \Lambda} v_F} l. \nn
\eea
There is no backreaction from the short-ranged interaction to the gauge fluctuation at one-loop order.
Thus, in the presence of fluctuating gauge bosons, the RG equation becomes
\bea\label{RG-u}
	\frac{du}{dl} = -\frac12 u+ \sqrt 2 \alpha_0 u^2+ \Big(\frac{\alpha_0}3+\frac38 \Big) gu+ \frac{\alpha_0}{2\sqrt 2} g^2.
\eea
In the RG equations, there are four fixed points in the $(u,g)$-plane, including the Gaussian fixed point $(0,0)$, the density-wave transition point $(\frac1{2\sqrt{2}\alpha_0},0)$ in the absence of gauge bosons, and two new fixed points emerging from the interplay between gauge fluctuations and short-ranged interactions: $
	 FP_\text{CFL} = \Big( \frac{6-(9+8 \alpha_0)\epsilon  - \sqrt{C(\epsilon)}}{24\sqrt{2} \alpha_0}, 2\epsilon \Big) \approx (2\sqrt2 \alpha_0 \epsilon^2, 2\epsilon ) $ and $
	 FP_\text{T} = \Big( \frac{6-(9+8 \alpha_0)\epsilon  + \sqrt{C(\epsilon)}}{24\sqrt{2} \alpha_0}, 2\epsilon \Big)$,
where $C(\epsilon)= (81+144 \alpha_0 - 1088 \alpha_0^2) \epsilon^2 - 12(9+8 \alpha_0) \epsilon + 36$ is a quadratic function in $\epsilon$. When $0< \epsilon < \epsilon_c$, $C(\epsilon)>0$, all of the four fixed points are physically accessible, and $FP_\text{CFL}$ ($FP_\text{T}$) corresponds to the CFL fixed point (density-wave transition point). Here $\epsilon_c \equiv \frac{6(9-8(3\sqrt 2 -1)\alpha_0)}{81+144 \alpha_0 -1088 \alpha_0^2}$ is a positive number. When $\epsilon<\epsilon_c$ the blue points in Fig.~\ref{u-g1} correspond to the Gaussian and the density-wave transition point without gauge fluctuation, while the red points correspond to $FP_\text{CFL}$ and $FP_\text{T}$. 

We also note that, in the presence of gauge fluctuations, the critical coupling strength of the $2k_F$ density-wave transition is significantly reduced.
More exotically, when $ \epsilon = \epsilon_c$, $C(\epsilon_c)=0$, the CFL fixed point and the transition point collide with each other, as shown in Fig.~\ref{u-g2}. The CFL transition fixed point is unstable against $2k_F$ density-wave instability. We would like to point out that such a fixed point collision is also found in previous literatures~\cite{son2009,herbut2014, Zhang2018}. When $\epsilon > \epsilon_c$, the CFL is totally preempted by density-wave orders as shown in Fig.~\ref{u-g3}. These results indicate that the NFL fixed point is unstable if the gauge fluctuation is strongly enough. We also note that the $FP_\text{CFL}$ is well controlled by $\epsilon$ expansion as can be seen from $FP_\text{CFL} \approx (2\sqrt2 \alpha_0 \epsilon^2, 2\epsilon )$ at small $\epsilon$. Moreover, at the merging point $\epsilon_c$, the values of two fixed points $FP_\text{CFL}$ and $FP_\text{T}$ are coincident, and so are controlled provided $\epsilon_c$ to be small. In the one-loop calculation, $\epsilon_c \approx 0.32 < 1$, indicates the scenario of the fixed-point collision is under control.

\section{Discussions} 
The large portion of CFL in the phase diagram in Fig.~\ref{Fig1} suggests $\epsilon<\epsilon_c$.
Although it is unclear how the bare interaction strengths, namely, the gauge coupling and the short-ranged interaction, change with the mixed form factors, the RG analysis is able to predict the wavevector of the density wave in the presence of the mass anisotropy. This is because the bare gauge coupling is enhanced by the mass anisotropy through the Fermi velocity.
Assuming $(u,g)=(u_0,g_0)$ for the isotropic CFL, we have $(u,g)=(u_0, \tilde\alpha^{1/4} g_0)$ at the patches ${\bf k}= (\pm \sqrt{2\tilde m_x \mu}, 0)$ of the anisotropic Fermi surface, where $\tilde \alpha \equiv \tilde m_x / \tilde m_y$ denotes the mass anisotropy of CFs (we will consider $\tilde\alpha\ge1$, since the opposite case is equivalent).  $\tilde \alpha$ is related to the mass anisotropy of electrons $\alpha$ through $\tilde \alpha =\sqrt{\alpha}$~\cite{Bhatt2017}.
It is easy to see that $\tilde\alpha^{1/4} g_0$ is the largest bare value in the elliptic Fermi surface, therefore, the above RG analysis predicts that the $2k_F$ instability occurs at $ 2k_F =2\sqrt{2 \tilde m_x \mu}$, which connects the Fermi points with the smallest Fermi velocity. This observation is consistent with $N(\mathbf{q})$ in Fig.~\ref{Fig:Nq}(c) near the transition point.
Note that this is a gauge fluctuation induced stripe transition.

Deep inside the charge density wave phase, we also find the switch of stripe orientations, as shown in Figs.~\ref{Fig:Nq}(c-d). This phenomenon might be beyond the CFL physics since it is further away from the critical point, however, it can be attributed to the reduction of Hartree energy cost when the stripe orientation coincides with the direction of the smaller mass~\cite{ZZ2017}. Moreover, from our ED results, the energy level crossing [see Fig.~\ref{Fig1}(b)] and the sudden jump in the first order derivatives [see Fig.~\ref{Fig1}(c)] suggest the transition from CFL to charge-density wave might be first order. We should note that it is still under debate whether the $2k_F$ density-wave transition is continuous. While Altshuler et al.~\cite{Altshuler1995} argues that a first-order transition occurs due to the strong $2k_F$ fluctuation at low energies, a more recent article by Sykora et al.~\cite{Metzner2018} shows a second-order transition is also possible. It will also be an excellent task to investigate the critical phenomena in $2k_F$ transitions of NFL, which we leave for future works.

Here, the RG analysis does not rely on the particle-hole symmetry. Therefore, the density-wave instability should be possible to extend to CFL at other filling factors, such as $\nu=1/4$.  A recent numerical study~\cite{ZZ} suggests that, for $\nu=1/4$ with mass anisotropy, the CFL state in $n=0$LL is robust while a stripe-like phase emerges in $n=1$LL, then one can expect that tuning the mixed form factor would trigger the density-wave instability of CFL at $\nu=1/4$ by the same mechanism proposed here.

The experimental probe of the various instabilities of CFL still presents many challenges and under intensive investigations. Previous studies mainly focus on detecting the pairing instability of CFL, which has been proposed by tuning the subband level crossings~\cite{Shayegan2011,Smet2012} or applying hydrostatic pressure~\cite{Csathy2016,Csathy2017,Csathy2018} in GaAs quantum wells, or by tuning either the perpendicular magnetic field or the interlayer electric bias in bilayer graphene~\cite{Young2017a,Young2017b,ZZ2019}. In particular,  the hydrostatic pressure experiments~\cite{Csathy2016,Csathy2017,Csathy2018} have found that tuning the pressure through $P_{c1}$ would trigger the transition from MR to an anisotropic compressible phase, which is consistent with either a stripe phase~\cite{Rezayi1999,HR2000,ZZ2017} or nematic phase~\cite{Kim2018} . Interestingly, further increasing the pressure to $P_{c2}$ leads to a transition to an isotropic compressible phase, which might be relevant to the density-wave instability, particularly considering that the pressure is believed to change the LL mixing parameters~\cite{Csathy2018}. However, we should also note the pressure-driven platform is hard to be captured by an ideal Hamiltonian microscopically, which would be an interesting direction for a future study. Moreover, the mixed form factor could be realized and tunable in bilayer graphene by an interlayer electric bias and magnetic field~\cite{Papic2011, Young2017a,Young2017b,ZZ2019},  then breaking the rotational symmetry may potentially probe the density-wave instability of CFL. The mass anisotropy exists in AlAs quantum wells~\cite{Shayegan2006,Gokmen2010} in nature or could be introduced by applying an in-plane field~\cite{ExpBx} or uniaxial strain~\cite{Rokhinson2011,Jo2017}, so then realizing a density-wave instability on top of an anisotropic CFL is also a promising direction to pursue experimentally.

\section*{Acknowledgements}

We thank Lukas Janssen, Sung-Sik Lee, D. N. Sheng, Inti Sodemann and Brian Swingle for helpful discussions. S.-K.J. is supported by the Simons Foundation via the It From Qubit Collaboration. We are grateful to D. N. Sheng for kindly providing some computational resources to finish some numerical calculations in this work. 
The part of this work carried out at Harvard was supported by the funding via Ashvin Vishwanath. The part of this work carried out at KITS was supported by the Fundamental Research Funds for the Central Universities, the start-up funding of KITS at UCAS, and the Strategic Priority Research Program of CAS (Grant No. XDB33000000).

Shao-Kai Jian and Zheng Zhu contributed equally to this work.

\begin{widetext}
\appendix
\section{Composite fermi liquid} \label{append:pair}
The action of two patches is given by
\bea
	S &=& S_f + S_a + S_{\text{int}} \\
	S_f &=& \sum_s \int d^3 x \psi_s^\dag (\partial_\tau - i s v_F \partial_x - \frac1{2K} \partial_y^2 ) \psi_s  \\
	S_a &=& \int \frac{d^3k}{(2\pi)^3} |k_y|^{1+\epsilon} |a(k)|^2 \\
	S_{\text{int}} &=& \sum_s  \int d^3 x  \frac{s e }{\sqrt N} a \psi_s^\dag \psi_s
\eea
where $s= \pm$ denotes the two patches, $\psi_s$ and $a$ refer to composite fermion and emergent gauge field, respectively. $v_F$ and $K$ capture the fermi velocity and curvature of the patch, and $e$ is the Yukawa coupling between fermion and gauge boson. The above action is believed to describe various interesting systems, such as $U(1)$ quantum spin liquids with a large spinor fermi surface and composite fermi liquids in the half-filled Landau level. Here, we mainly focus on the latter case, and the above action is a patch description of the Halperin-Lee-Read (HLR) theory~\cite{Halperin1993}. The summation over $N$ flavors of the patch fermion is implicit, and $\epsilon$ is the expansion parameter. $\epsilon = 0$ corresponds to the long-range Coulomb interaction~\cite{Senthil2010, Kim2017}. In the noninteracting limit, the action is invariant under scaling transformation dictated by the scaling dimensions,
\bea
	[k_x] = 1, \quad [k_y]= \frac12, \quad [\omega]= 1, \quad [\psi] = \frac34, \quad [a] =1- \frac\epsilon4, \quad [e]= \frac\epsilon4.
\eea
The RG calculation is controllable in the large-$N$ and small $\epsilon \sim \frac1N$ expansion~\cite{Senthil2010}. The patch theory is an effective description in the range $ |k_x|, k_y^2 < \Lambda$. In the following, we integrate out the high energy mode, $ \sqrt {\Lambda e^{-l}} < |k_y| < \sqrt{\Lambda}$ to generate RG equations, where $l>0$ is the running parameter. There is no renormalization to boson propagators because it is nonlocal. The rationale for using a nonlocal bare kinetic term for the gauge boson lies in the fact that the boson kinetic potential does not receive corrections up to three-loop~\cite{Sachdev2010}. The fermion self-energy is (Fig.~\ref{selfEnergy})
\bea
	\Sigma_s(p) &=& - \frac{e^2}N \int \frac{d^3 k}{(2\pi)^3} D(k) G_s(k+p) = - \frac{e^2}{N} \int \frac{d^3 k}{(2\pi)^3} \frac1{|k_y|^{1+\epsilon}} \frac1{-i(k_0+ p_0) + s v_F(k_x + p_x) + \frac1{2K}(k_y + p_y)^2} \\
	&=& - \frac{e^2}{N} \int \frac{d^3 k}{(2\pi)^3} \frac{i\pi \text{sgn}(k_0+p_0) \delta(s v_F(k_x + p_x) + \frac1{2K}(k_y + p_y)^2)}{|k_y|^{1+\epsilon}} \\
	&=& - \frac{i e^2}{2(2\pi)^2 Nv_F} \int dk_y d k_0\frac{ \text{sgn}(k_0+p_0)}{|k_y|^{1+\epsilon}} =
	 - \frac{i e^2}{2\pi^2 Nv_F} p_0 \int_{\sqrt{\Lambda e^{-l}}}^{\sqrt \Lambda} dk_y \frac{ 1}{|k_y|^{1+\epsilon}} \approx - \frac{i e^2}{4 \pi^2 Nv_F} p_0 l,
\eea

\begin{figure}
	\centering
	\subfigure[]{\label{selfEnergy}
		\includegraphics[height=1.4cm]{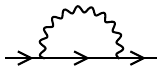}}~~
	\subfigure[]{\label{vertex3}
		\includegraphics[height=2.cm]{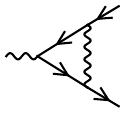}}
	\caption{The Feynman diagrams: (a) fermion self-energy, and (b) fermion-boson vertex.}
\end{figure}
The vertex correction is (Fig.~\ref{vertex3})
\bea
\Gamma_3 &=& \sum_s \int  \frac{d^3 k}{(2\pi)^3}  G_s^2(k) D(k) = \sum_s \int  \frac{d^3 k}{(2\pi)^3}  \frac1{|k_y|^{1+\epsilon}} \frac1{[-i(k_0+ p_0) + s v_F(k_x + p_x) + \frac1{2K}(k_y + p_y)^2]^2},
\eea
which vanishes because the poles of $k_0$ lie in the same plane. In terms of the dimensionless coupling constant $g \equiv \frac{e^2}{\pi^2 v_F \Lambda^{\epsilon/2}}$ that captures the effective Yukawa coupling, we have following RG equations,
\bea
	\frac{d g}{dl} &=& \frac{\epsilon}2 g - \frac{g^2}{4}.
\eea
The presence of a nontrivial stable fixed point $g^*=2 \epsilon$ corresponds to the non-fermi liquid (NFL) interacting strongly with the gauge field.

\section{Cooper instability and $2k_F$ density-wave instability} \label{append:RG}
\begin{figure}
	\centering
	\subfigure[]{\label{bcs1}
		\includegraphics[height=1.6cm]{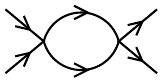}}~~
	\subfigure[]{\label{bcs2}
		\includegraphics[height=1.6cm]{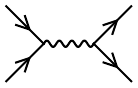}}
	\caption{The Feynman diagrams in the particle-particle channel. (a) is the one-loop corrections of from the four-fermion BCS interaction. (b) is the interpatch interaction resulting from integrating out high energy gauge flucuation.}
\end{figure}
Despite the long-range interactions between the composite fermion mediated by the gauge field, there are local interactions between the composite fermion that might generate pairing or stripe instability. Thanks to the Pauli exclusion principle of fermions, among infinite channels of four-fermion interactions only BCS and the forward-scattering channel survive in the low energy~\cite{Shankar1994}. For simplicity, we will send $N=1$ in the following and consider four-fermion interactions. We first consider the BCS Hamiltonian for the nondegenerate fermi surface,
\bea
	H_{\text{BCS}} = -\int \frac{d^2 k}{(2\pi)^2} \frac{d^2 k'}{(2\pi)^2} V({\bf k, k'}) \psi^\dag({\bf k}) \psi^\dag(- {\bf k}) \psi( {\bf k'}) \psi(- {\bf k'}),
\eea
where $\psi$ denotes the fermi surface, and $V$ is the strength of the BCS interaction. It is well known that the pairing instability is marginally relevant for fermi liquids~\cite{Shankar1994}. The RG equation is given by (Fig.~\ref{bcs1}, and we consider spherical fermi surface for simplicity),
\bea
	\frac{d v_j}{dl} = -  v_j^2,
\eea
where $v_j= \frac{k_F}{2\pi v_f} V_j$ and $V_j = \int \frac{d\theta}{2\pi} V(\theta) e^{i \theta j}$.  Different from fermi liquids, the presence of an emergent gauge boson in a composite fermi liquid suppress the pairing instability. Indeed, as shown in Fig.~\ref{bcs2}, integrating out the high-energy mode of gauge fluctuation will generate an interpatch interaction~\cite{Senthil2015}. Here we review the calculations~\cite{Senthil2015}. For a small-angle BCS interaction, $ \theta \sim 0$, we have the correction from the gauge fluctuation,
\bea
	\delta V({\bf k_1, k_2}) = \frac{e^2}{2N} D(\bf k_1 - k_2),
\eea
while it also contributes to $V(\theta \sim \pi)$. Taking both of these into consideration, the corrections to the BCS interaction are
\bea
\delta v_j &=& \frac{k_F}{2\pi v_f} \int \frac{d\theta}{2\pi} \delta V(\theta) e^{i \theta j} \approx \frac{1}{\pi v_f} \frac{e^2}{N} \int_{\sqrt{\Lambda e^{-l}}}^{\sqrt \Lambda} \frac{d k}{2\pi} \frac1{k^{1+\epsilon}} \approx \frac{e^2}{4\pi^2 v_f N} l.
\eea
Therefore, including the gauge fluctuation, the RG equation reads
\bea
	\frac{d v}{dl} = -v^2 + \frac{g}{4}.
\eea
Because of the suppression from gauge fluctuation, the BCS instability is no longer marginally relevant. Instead, it requires a finite bare BCS interaction to drive the composite fermi liquid into the paired state. Note that in the context of the half-filled Landau level, for example, in the $\nu=5/2$ filling fraction, the system favors $p+i p$ pairing, which is the famous Moore-Read Pfaffian state~\cite{Moore1991, Read2000}.

On the other hand, we consider the four-fermion interaction within the patch theory in the following,
\bea
	S &=& S_f + S_a + S_{\text{int}} + S_4 , \\
	S_4 &=& U \int d^3 x \psi_+^\dag \psi_+ \psi_-^\dag \psi_- .
\eea
In the patch theory, the four-body interaction is irrelevant, which is consistent with the fact that forward-scattering does not affect the existence of a fermi surface, and the perturbative calculation should be valid. The correction reads
\bea
	\Gamma^{(a)}_4 &=& - 2U^2 \int \frac{d^3 k}{(2\pi)^3} G_+(k) G_-(k)= \frac{2\sqrt{2K} U^2}{v_F} \int \frac{d^3 q}{(2\pi)^3} \frac1{q_0+i (q_x + q_y^2)} \frac1{q_0 - i (q_x - q_y^2)} \\
	&=& \frac{4\sqrt{2K} U^2}{(2\pi)^2 v_F}  \int_{\sqrt{\Lambda e^{-l}}}^{\sqrt \Lambda} d q_y \int^\infty_{q_y^2} d q_x\frac{e^{-q_x^2 /\Lambda^2}}{q_x} \approx  \frac{\sqrt 2 \Gamma(0,1)}{ \pi^2} \frac{\sqrt{\Lambda K} U^2}{v_F} l ,
\eea
where $\Gamma(n,x) \equiv \int_x^\infty dt t^{n-1} e^{-t} $ is the incomplete Gamma function, and $\Gamma(0,1) \approx 0.219$. In the calculation, we have introduced a regularization function $e^{-q_x^2/\Lambda^2}$ to regularize the UV divergence. Without gauge fluctuation, the RG equation of the dimensionless coupling constant $u \equiv \frac{\sqrt{K\Lambda}}{\pi^2 v_F} U$ is
\bea
	\frac{du}{dl}= - \frac12 u+ \sqrt2\Gamma(0,1) u^2,
\eea
which shows that an instability only occurs at a finite interaction strength, and the fermi liquid is perturbatively stable.

The corrections from gauge fluctuation read
\bea
	\Gamma^{(b)}_4 &=& - \frac{2e^2}{3} \int \frac{d^3 k}{(2\pi)^3} G_R(k) G_L(k) D(k) = \frac{2\sqrt{2K} e^2 u}{3 v_F} \int \frac{d^3 q}{(2\pi)^3} \frac1{q_0+ i(q_x + q_y^2 )}\frac1{q_0- i(q_x - q_y^2 )} \frac1{|\sqrt{2K} q_y|^{1+\epsilon}} \\
	&=& \frac{4e^2 u}{3(2\pi)^2 v_F} \int_{\sqrt{\Lambda e^{-l}}}^{\sqrt \Lambda} \frac{d q_y }{|q_y|^{1+\epsilon}} \int^\infty_{q_y^2} d q_x\frac{e^{-q_x^2 /\Lambda^2}}{q_x}  \approx \frac{\Gamma(0,1)}{3\pi^2} \frac{e^2 u}{v_F} l,
\eea
and
\bea
	\Gamma^{(c)}_4 &= & -e^4 \int \frac{d^3 k}{(2\pi)^3} G_R(k) G_L(k) D^2(k)= \frac{\sqrt{2K} e^4}{v_F} \int \frac{d^3 q}{(2\pi)^3} \frac1{q_0+ i(q_x + q_y^2 )}\frac1{q_0- i(q_x - q_y^2 )} \frac1{|\sqrt{2K} q_y|^{2(1+\epsilon)}} \\
	&=& \frac{2e^4}{(2\pi)^2 N^2 \sqrt{2K} v_F} \int_{\sqrt{\Lambda e^{-l}}}^{\sqrt \Lambda} \frac{d q_y }{|q_y|^{2(1+\epsilon)}} \int^\infty_{q_y^2} d q_x\frac{e^{-q_x^2 /\Lambda^2}}{q_x} \approx \frac{\Gamma(0,1)}{2\sqrt2 \pi^2 N^2} \frac{e^4}{\sqrt{K \Lambda} v_F} l.
\eea

These corrections lead to the RG equations in the main text.

%----------------------------------------------------------------------------------------------------------------------------------------------------------------------------------------
\begin{figure*}[tbp]
\begin{center}
\includegraphics[width=0.9\textwidth]{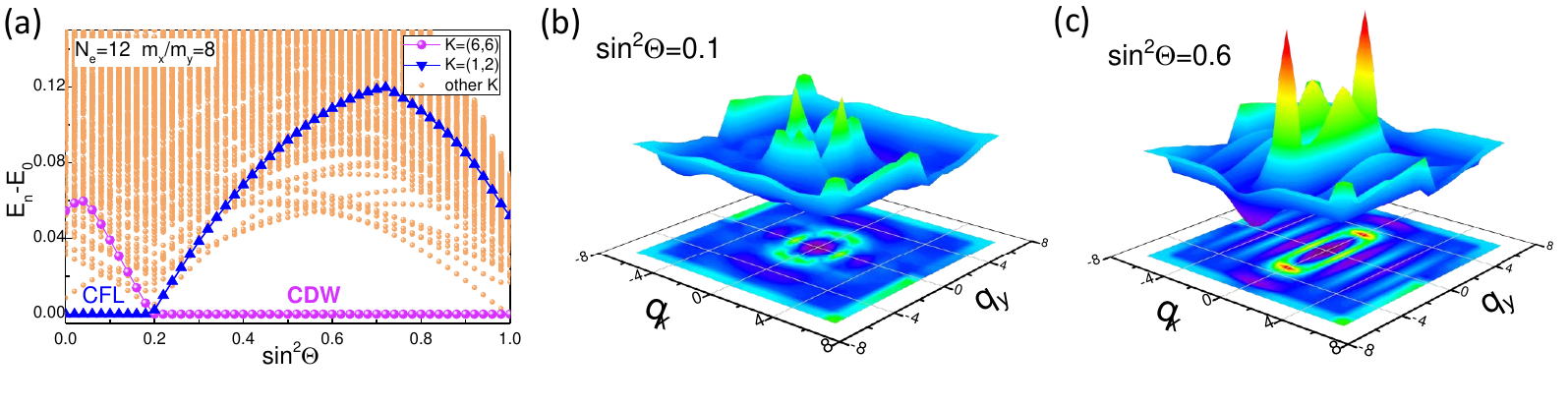}
\end{center}
\par
\renewcommand{\figurename}{Fig.}
\caption{(Color online) (a)The evolution of energy spectra as a function of  $\text{sin}^2\Theta$ for a half-filled Landau level with $N_e=12$ electrons  and mass ratio $m_y/m_x=8$.  The nature of the different phases can be identified from the static structure factor $N(\mathbf{q})$ of the density-density correlation.  Panels (b-c) show $N(\mathbf{q})$ in the CFL phase with  $\text{sin}^2\Theta=0.1$ (a) and in the charge density wave phase with $\text{sin}^2\Theta=0.6$ (c).}
\label{FigS_N12}
\end{figure*}
%----------------------------------------------------------------------------------------------------------------------------------------------------------------------------------------
\begin{figure*}[tbp]
\begin{center}
\includegraphics[width=0.9\textwidth]{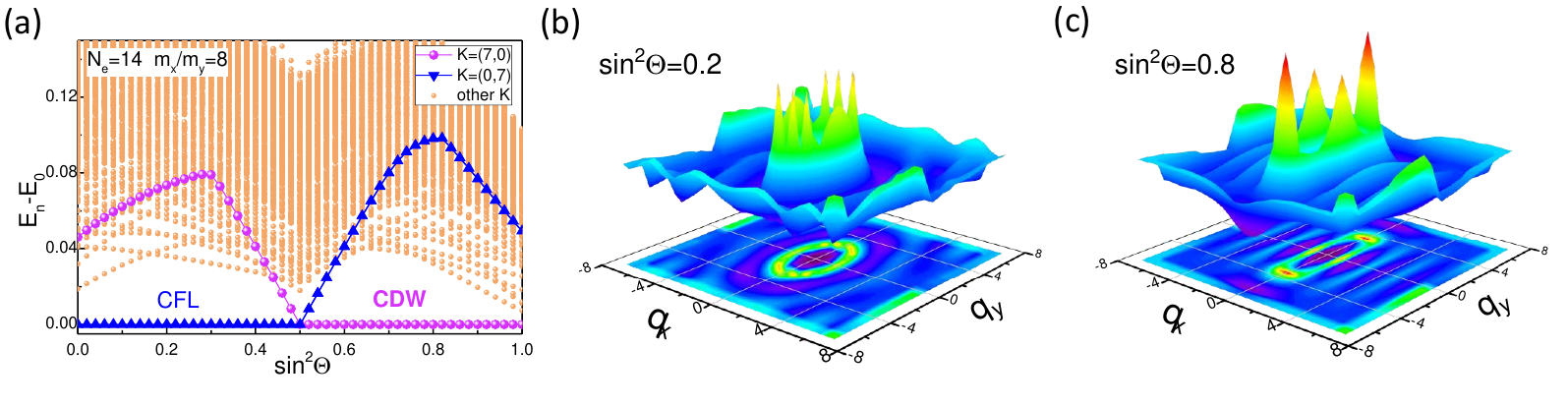}
\end{center}
\par
\renewcommand{\figurename}{Fig.}
\caption{(Color online) (a)The evolution of energy spectra as a function of  $\text{sin}^2\Theta$ for a half-filled Landau level with $N_e=14$ electrons  and mass ratio $m_y/m_x=8$.  The nature of the different phases can be identified from the static structure factor $N(\mathbf{q})$ of the density-density correlation.  Panels (b-c) show $N(\mathbf{q})$ in the CFL phase with  $\text{sin}^2\Theta=0.2$ (a) and in the charge density wave phase with $\text{sin}^2\Theta=0.8$ (c).}
\label{FigS_N14}
\end{figure*}

\section{Finite Size Effect} \label{append:size}

In the main text, we mainly show the results for $N_e=16$ systems, but we also have checked the other system and found similar results, as shown in Fig.~\ref{FigS_N12} and Fig.~\ref{FigS_N14} for $N_e=12$ and $N_e=14$ systems. The CFL to CDW transition can be identified from the evolution of energy spectra as a function of  $\text{sin}^2\Theta$ [see Fig.~\ref{FigS_N12}(a) and Fig.~\ref{FigS_N14} (a) for mass ratio $m_y/m_x=8$]. The nature of the CFL phase and CDW phase can be identified from the static structure factor $N(\mathbf{q})$ of the density-density correlation. For the CFL phase, $N(\mathbf{q})$ exhibits a strong $2k_F$ scattering feature induced by the scattering among CFs close to the Fermi surface, as shown in Fig.~\ref{FigS_N12}~(b) and Fig.~\ref{FigS_N14}~(b). For the CDW phase, as shown in Fig.~\ref{FigS_N12}~(c) and Fig.~\ref{FigS_N14}~(c), there are two sharp peaks in $N(\mathbf{q})$ in the same direction, which can be regarded as the hallmark of charge ordering with the wave vector determined by the position of the peaks.

\end{widetext}

\end{document}